\documentclass[fleqn,10pt]{wlscirep}
\usepackage[utf8]{inputenc}
\usepackage[T1]{fontenc}
\usepackage{amsmath,amsfonts,amsthm,amssymb,algorithm,algpseudocode}

\newcommand{\CatWeightij}{$\widetilde{\rho}_{ij,k}$}
\newcommand{\CatWeightj}{$\widetilde{\rho}_{.k}$}

\title{Comparing multiple networks using the Co-expression Differential Network Analysis (CoDiNA)}

\author[1,2,*]{Deisy Morselli Gysi}
\author[3]{Tiago de Miranda Fragoso}
\author[4]{Volker Buskamp}
\author[5,6]{Eivind Almaas}
\author[7,*]{Katja Nowick}

\affil[1] {Department of Computer Science, Interdisciplinary Centre of Bioinformatics, University of Leipzig, Leipzig, D-04109, Leipzig.}
\affil[2] {Swarm Intelligence and Complex Systems Group, Faculty of Mathematics and Computer Science, University of Leipzig,  Leipzig, D-04109, Leipzig.}
\affil[3] {Funda\c{c}\~ao Cesgranrio, Rio de Janeiro, 20261-903, Brazil.}
\affil[4] {Technische Universität Dresden, DFG Research Center for Regenerative Therapies, 01307 Dresden, Germany}
\affil[5] {Department of Biotechnology,  NTNU - Norwegian University of Science and Technology, Trondheim, Norway.}
\affil[6] {K.G. Jebsen Centre for Genetic Epidemiology, NTNU - Norwegian University of Science and Technology, Trondheim, Norway.}
\affil[7] {Human Biology Group, Institute for Biology, Department of Biology, Chemistry, Pharmacy, Freie Universitaet Berlin, Koenigin-Luise-Str. 1-3, D-14195 Berlin, Germany.}

\affil[*]{ corresponding authors: deisy@bioinf.uni-leipzig.de, katja.nowick@fu-berlin.de}

\algrenewcommand\algorithmicrequire{\textbf{Input:}}
\algrenewcommand\algorithmicensure{\textbf{Output:}}

\begin{abstract}
	Biomedical sciences are increasingly recognising the relevance of gene co-expression-networks for analysing complex-systems, phenotypes or diseases. When the goal is investigating complex-phenotypes under varying conditions, it comes naturally to employ comparative network methods. While approaches for comparing two networks exist, this is not the case for multiple networks.
	Here we present a method for the systematic comparison of an unlimited number of networks: \textbf{Co}-expression \textbf{Di}fferential \textbf{N}etwork \textbf{A}nalysis (CoDiNA) for detecting links and nodes that are common, specific or different to the networks. 
	Applying CoDiNA to a neurogenesis study identified genes for neuron differentiation. Experimentally overexpressing one candidate resulted in significant disturbance in the underlying neurogenesis' gene regulatory network.
	We compared data from adults and children with active tuberculosis to test for signatures of HIV. We also identified common and distinct network features for particular cancer types with CoDiNA. These studies show that CoDiNA successfully detects genes associated with the diseases.
\end{abstract}
\begin{document}
	
	\flushbottom
	\maketitle
	%
	%
	\thispagestyle{empty}

	\section*{Main}
	Complex systems, exemplified by biological pathways, social interactions, and financial markets, can be expressed and analysed as systems of multi-component interactions\cite{Kuntal2016}. In systems biology, it is necessary to develop a thorough understanding of the interactions between factors, such as genes or proteins.  Gene co-expression networks have been especially effective in identifying those interactions\cite{barabasi2004network, bansal2007infer, furlong2013human,  dempsey2013mining}. In gene co-expression networks,  nodes represent genes and a weighted link between a pair of genes represents their connection, often calculated as a correlation. 
	The sign of the relation may suggest an up- or down-regulation of one factor by the other\cite{van2017gene}. It has been shown that different conditions have different underlying regulatory patterns and will therefore lead to different networks even for a single system\cite{berto2016consensus, Kuntal2016, gysi2017wto}. 
	
	While the analysis of expression differences allows for identification of genes that are significantly differentially  expressed between two or more conditions\cite{fukushima2013diffcorr, furlong2013human, del2010diseases, bandyopadhyay2010rewiring, de2010differential}, it does not enable the  investigation of more complex patterns, such as changes in the rewiring of the regulatory relationships of genes. Differential network analyses are able to capture changes in gene relationships and are thus exceptionally suitable for understanding complex phenotypes and diseases\cite{furlong2013human}.
	
	In order to distinguish similarities and differences between system-level activity patterns, it has become common practice to compare multiple networks, derived from different conditions. 
	This is usually followed by an analysis of links that are shared between all networks or are chiefly specific to some of the networks. 
	Classification of links according to the concepts of being present, different or absent in some networks is essential to understanding how different environments influence particular interactions. 
	Several methods for such comparisons exist for pairs of networks\cite{Chen2004, Liang2006, watson2006coxpress, Tian2007, wu2008integrated, Dost2008, Lena2013, bass2013using, fukushima2013diffcorr, voigt2017composite}. However, in many situations it is of great interest to compare more than just two networks simultaneously. Such a set of networks could e.g. be gene-regulatory networks arising from different species, tissues or diseases, or co-existence networks from different environments. 
	For instance, an evolutionary study conducted a pairwise comparison between humans, chimpanzees and rhesus macaques to pinpoint similarities and differences in the prefrontal cortex \cite{berto2018}. 
	In a recent medical study,  the authors  compared enriched gene functions using the Gene Ontology \cite{yang2014gene} instead of comparing the networks of the multiple cancers. Another study generated a network that involved only differentially expressed genes  \cite{yang2017identification} extracted from their multiple networks.  These studies could have  profited extensively from applying a method capable of systematically comparing multiple networks simultaneously.
	
	Kuntal and collaborators\cite{Kuntal2016} proposed a method, CompNet, that may address the comparison of multiple networks. However, the focus of CompNet is on the visualisation of the union, intersections and exclusive links of the analysed networks.
	Thus, a method that is capable of comparing both links and nodes of any number of networks is still lacking.
	Here, we propose a novel method for that purpose, CoDiNA (\textbf{Co}-expression {\bf Di}fferencial {\bf N}etwork {\bf A}nalysis for n Dimensions), implemented as an \texttt{R} package.
	This package also includes an interactive tool for network visualisation.
	Our method was first applied to find common, specific or different links and nodes in a study of neurogenesis of induced Pluripotent Stem Cells (iPSC) with or without the presence of the micro RNA miR-124\cite{}. 
	CoDiNA identified modules of genes that were present in miR-124 knockout or wildtype cells and a hub-gene for the time point of highest network differences. Overexpressing this hub gene during neurogenesis resulted in a clear repression of neuronal differentiation. 
	Here we further demonstrate the power and versatility of our method with two example applications. In one example we compare expression data from tuberculosis (TB) patients with or without HIV.  The co-infection of HIV and TB is of strong medical importance since, to this date, it is challenging to detect TB in the presence of an HIV infection. In the other showcase, we compare three types of cancers to understand which molecular signatures they have in common and how they differ.
	
	\section*{Results and Discussion}
	
	To perform a comparison of co-expression networks, CoDiNA requires as input a set of networks to be assessed. The networks can be constructed using a correlation method, but should only contain links that are statistically significant given a predefined $p$-value threshold. We classify each link into one of three $\Phi$ categories based on its weight: a link is said to be an $\alpha$ link if it is present in all networks with the same sign, i.e., it is an interaction that is common to all networks.  
	A link is called $\beta$ if it is present in all networks but with different signs of the link's weight, i.e., it represents a different kind of interaction in at least one network. The biological interpretation of this category is that a particular gene changed its function, so that a gene that up-regulates another gene in one condition down-regulates the same gene in another condition (or vice versa).
	A link is considered a $\gamma$ link, if it is present in some networks but not all, regardless of the sign of the link's weight, i.e., this link is specific to at least one network. To further characterise \textit{how} a  particular link is different or specific, we are assigning a sub-category, which we refer to as $\widetilde{\Phi}$. This subcategory clarifies to which condition a link is specific or in which condition it has changed.
	
	In order to avoid false associations, i.e. the incorrect inference that a particular gene is associated with a specific condition, we require that all investigated {\it nodes} are present in all networks: If a node is absent in at least one network, we remove all of its links from the networks in which it is present. This does not apply to nodes that are present but have no significant links. In this case, it is assigned a (weight) value of zero, thus allowing all measured nodes to be included in the analysis, even when they only have not significant links.
	
	The weight value of the link between gene $i$ and $j$, denoted by $\rho_{ij}$, is defined in the interval $\left[-1,1\right]$. To denote links as positive, negative, or neutral we divide this interval into three equal parts. To compare networks whose intervals might vary, one option is to normalise the data inside the interval to $\left[-1,1\right]$; we refer to this approach as \textit{stretch}. This step is particularly important to compare networks that were not measured under the same experimental conditions. 
	For a set of $\mathbb{W}$ networks of $\mathbb{N}$ nodes, each one of the $\mathbb{E}$ links has a specific link weight $\rho_{ij,k}$ connecting nodes $i$ and $j$ in network $k$. Each link weight $\rho_{ij,k}$ is categorised into $1$, $0$ or $-1$ depending on its presence or absence in all, some, or none of the networks. If a particular link is absent in all of the $W$ networks, this link is removed from subsequent analyses. 
	If a link has a signal different than the link of the reference network, its $\widetilde{\Phi}$ category will be 
	$\beta_{N_1}$, where $N_1$ is a set listing the networks for which the signal changes (assuming the first network as the reference). Accordingly, if a link is specific, the category is
	$\gamma_{N_2}$, where $N_2$ is a set listing the networks for which the link exists.
	
\begin{figure}[ht]
		\centering
		\includegraphics[width=1\textwidth]{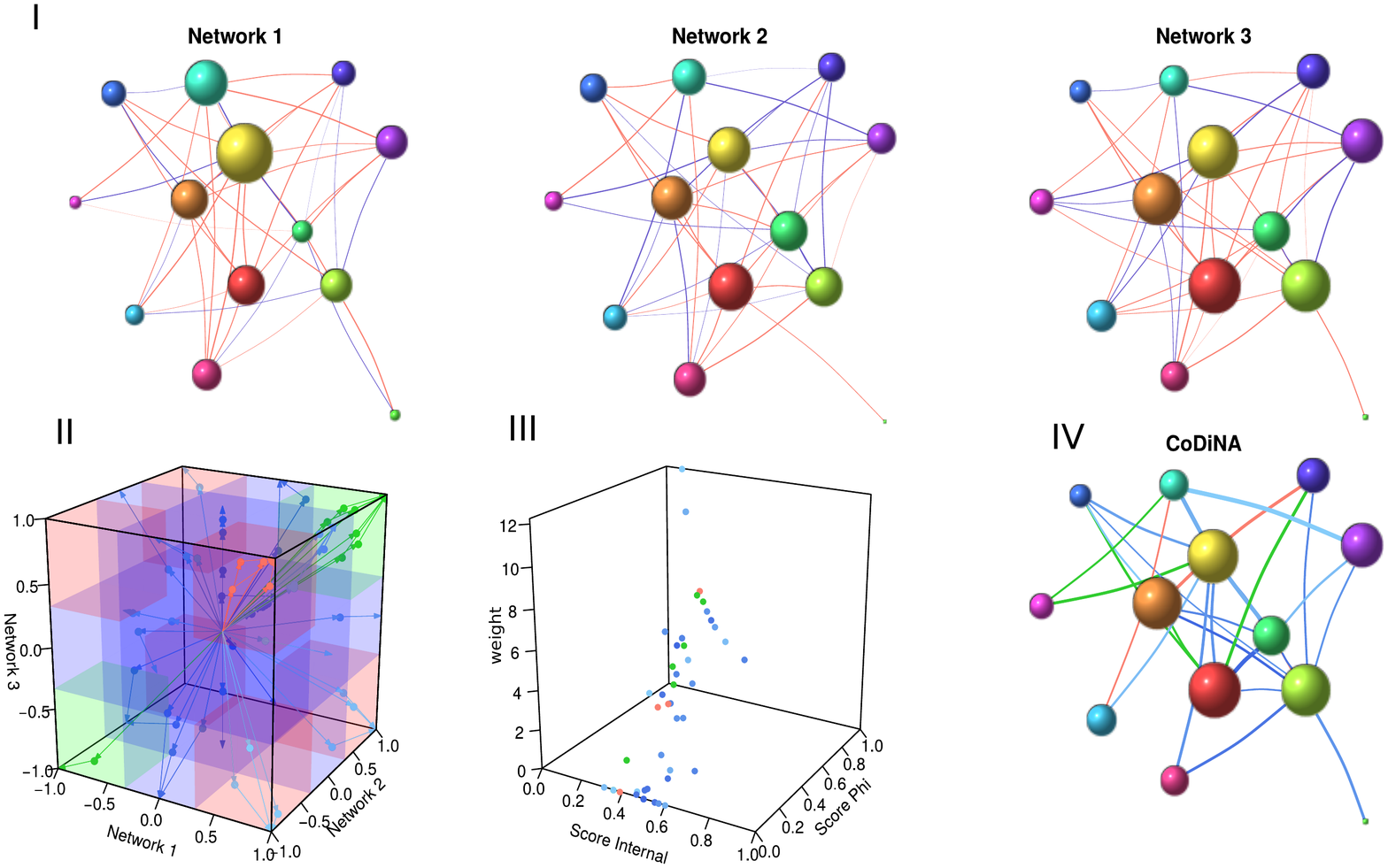}
		\caption{\textbf{Visual representation of the CoDiNA method for a $3$-network comparison.} {\bf Panel I} displays three independent networks to be compared. Red links represent positively correlated gene-pairs, and  blue links negatively correlated ones. Node-size is relative to node strength. \textbf{Panel II} shows the geometrical representation of CoDiNA:  a three-dimensional scatter-plot that is derived from plotting the weights from each link in the three networks. Different $\Phi$ regions: green for $\alpha$ ({Common}) links;  $\beta$ ({Different}) are red; blue corresponds to $\gamma$ ({Specific}) links. Scores are represented as arrows: arrows away from centre to the point correspond to $\widetilde{\Phi}$ scores ($\Delta_{\widetilde\Phi}$), while arrows from the point towards the centre represent the internal scores ($\Delta_{\widetilde{\rho}}$). {\bf Panel III} shows the relationship between the scores.  {\bf Panel IV} exhibits the CoDiNA network filtered for a score-ratio greater than unity. Link-width corresponds to link strength, and colour represents the $\widetilde{\Phi}$ class.  Note that, not all links appear in the filtered network. 
		}
		\label{fig:method}
	\end{figure}

	After defining the $\Phi$ categories, we score the links to identify those that are stronger. 
	To this end, for every node $i = 1,2,\ldots,N$, we interpret the array of link weights $\left(\rho_{1i},\ldots,\rho_{Wi}\right)$ as a point in a $W$-dimensional Euclidean space (Fig. \ref{fig:method}, panel II). 
	Thus, all points are contained in the cube determined by the Cartesian product $\left[-1,1\right]^W$.
	As such, a link that is closer to the centre of a $W$-dimensional cube is weaker than a link closer to the surface. This can be represented as $\Delta_{i,j}$, the Euclidean distance of links to the origin. To have representative values of all $\widetilde{\Phi}$, we normalise $\Delta_{i,j}$ by each $\Phi$ and $\widetilde{\Phi}$ category. 
	We denote the normalised distance as $\Delta^*_{i,j}$.  The  $\Delta^*_{i,j}$ normalised by $\widetilde{\Phi}$ is denoted as $\Delta_{\widetilde{\Phi}}$.
	We also calculate what we term the internal score, denoted as $\Delta_{\widetilde{\rho}}$, a categorical weight
	defined as the distance of each link to the best well-defined theoretical point that the link belongs to, $\widetilde{\rho}_{i,j}$. 
	This measure assures that a particular link is clustered in the correct $\widetilde{\Phi}$ category.  
	The ratio between $\Delta_{\widetilde{\Phi}}$ and $\Delta_{\widetilde{\rho}}$
	simultaneously scores how well clustered and strong a link is: The higher this ratio, the better defined a link is. Links with a ratio smaller than unity should be removed from posterior analysis.

	The classification of links is not sufficient to describe a network, which is why it is also necessary to categorise the nodes. We use a $\chi^2$ goodness-of-fit test to test if the frequency of the links in each category is different from the overall expected frequency in that category. 
	If the null hypothesis is rejected, the $\Phi$-category with the maximum number of links is assigned to that particular node. Similarly for the $\widetilde{\Phi}$ node categorisation. More detailed information on CoDiNA can be found in the Online Methods.
	
	\subsection*{Comparing CoDiNA to other methods}
	
	Evaluating multiple co-expression network methodologies for comparing networks is considerably challenging due to the lack of a gold standard network for multiple tissues\cite{Lichtblau2017}, of which all links are experimentally detected. Therefore, we are only able to identify similarities and differences among the methodologies. 
	
	To the best of our knowledge, only one other tool, CompNet\cite{Kuntal2016}, allows for the comparison of more than two networks. The focus of CompNet is on the visualisation of pairwise Jaccard-similarities from the union, intersections and exclusive links of those networks.
	It includes features such as pie-nodes and links to allow the user to identify  key elements of the network. Elements are identified by providing a distribution of global graph properties, such as the amount of nodes, links, density, clustering coefficient, average path length and diameter, of the networks.
	Even though, building a visualisation tool, is not the focus of CoDiNA, we also incorporated an interactive tool for visualisation of the final network, and CoDiNA provides summary statistics of the network, such as the total number of links and nodes as well as how many links and nodes have been classified as common, different or specific to each category. 
	
	Lichtblau et. al. (2017)\cite{Lichtblau2017} compared ten differential network analysis methods that are able to perform pairwise comparison. The authors split the methods into two main categories: Local search and Global search. Global methods focus on  changes in the network topology while local methods search for changes in the nodes. CoDiNA combines both: it first searches for changes in the topology of the networks and then for specificity of the nodes. This allows investigating both features with one tool. Changes in network topology indicate alterations in affected pathways or regulatory relationships, while changes concerning specific nodes can evaluate the importance of particular genes for the network and suggest genes that might be responsible for the topology differences. Together, the local and global changes are crucial for understanding the functional effects of network changes. 
	
	\subsection*{CoDiNA: an \texttt{R} package to compare co-expression networks}
	
\begin{figure}
		\centering
		\includegraphics[width=0.9\textwidth]{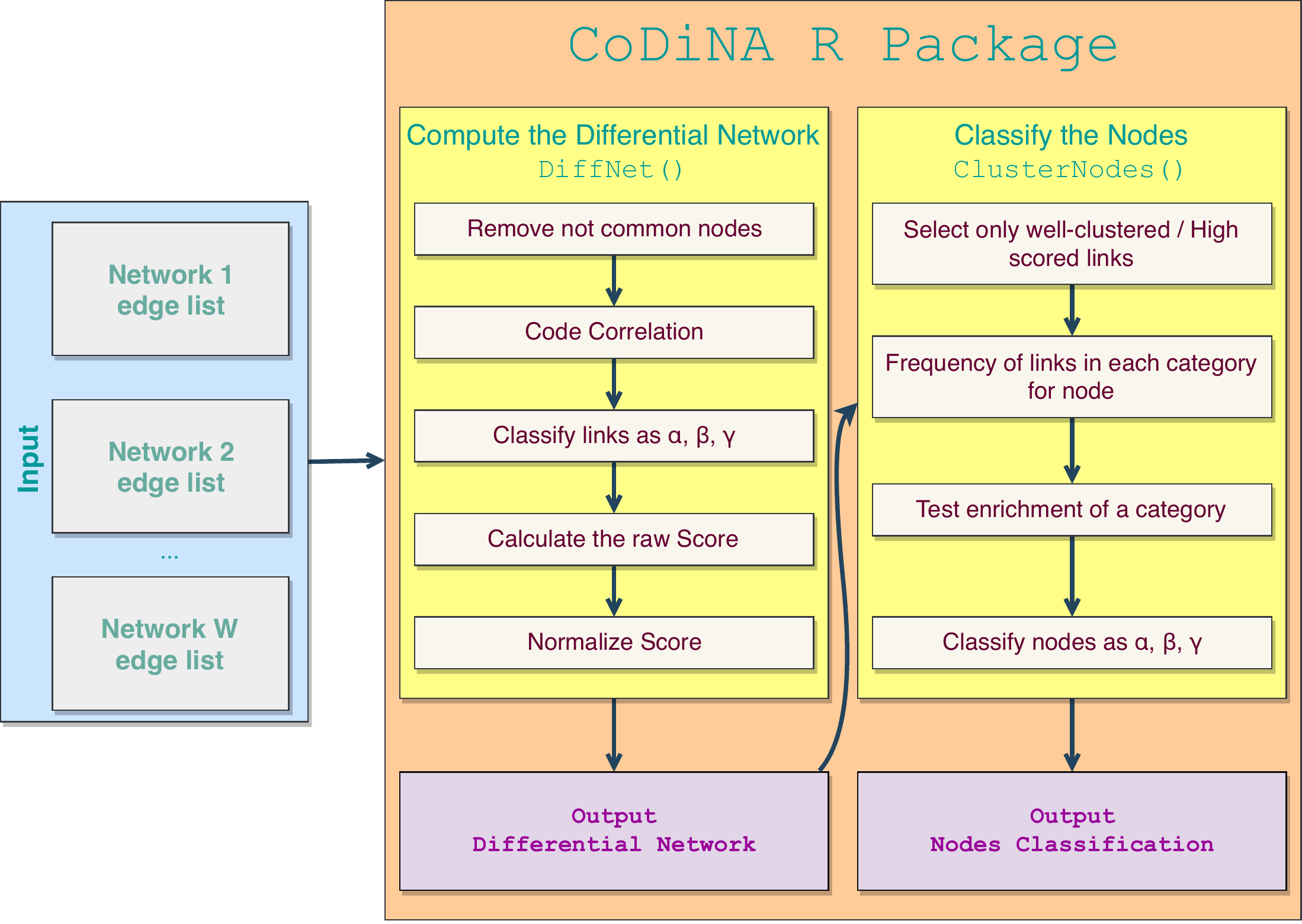}
		\caption{\textbf{Workflow process of the CoDiNA \texttt{R} package.}  Input data for the CoDiNA \texttt{R} package can be any network, filtered for only significant links.  Edge list is a list containing all the links and its weights. To links for which the $p$-value is not significant, the user can assign a weight of zero. 
			A minimum input requirement is of two networks with at least one link each. The function \texttt{MakeDiffNet()} clusters the links into the $\Phi$ and $\widetilde{\Phi}$ categories, calculates and normalises the scores. 
			Its output is used as input for clustering the nodes into categories by the function \texttt{ClusterNodes()}. 
			The \texttt{plot()} function can be used on the output from \texttt{MakeDiffNett()} and automatically calls the function \texttt{ClusterNodes()}.}
		\label{fig:Workflow}
	\end{figure}

	To make the proposed methodology publicly available, an \texttt{R} package called \texttt{CoDiNA} was developed (\ref{fig:Workflow}), where all the presented steps are implemented. The \texttt{R} package also includes an interactive visualisation tool. The main functions of the package are 
	\texttt{MakeDiffNet}, which categorises all the links into 
	$\Phi$ and $\widetilde{\Phi}$ categories, calculates the internal scores and the normalised scores, and \texttt{ClusterNodes}, which clusters the nodes into the different categories. The visualisation tool can be accessed using the function \texttt{plot}. 
	
	\section*{Pairwise version of CoDiNA: iPSC cells undergoing Neurogenesis had its process unravelled by CoDiNA}
	
	An earlier version of CoDiNA for a pairwise network comparison was already successfully applied to an expression dataset of human induced Pluripotent Stem Cells (iPSC)\cite{Kutsche2018} (\textit{in press}). The iPSCs were induced to undergo neuronal differentiation within four days. We compared expression patterns of wildtype and miR-124 deleted iPSCs over the time course of differentiation with the goal to uncover the function of miR-124 during neurogenesis. The experiment was conducted in seven replicates to facilitate the construction of a co-expression network for each day for the wildtype and  the knockout cells. Using the pairwise version of CoDiNA, we revealed strong network differences between wildtype and knockout cells on day $3$ of neuronal differentiation. For that day, CoDiNA classified the transcription factor (TF) ZNF787 as the TF with most specific links, suggesting it as one of the drivers of miR-124 induced network changes. Since ZNF787 is more highly expressed in miR-124 knock-out compared to wildtype iPSC, we overexpressed ZNF787 in the wildtype cells to experimentally validate our result. This overexpression still resulted in neuronal differentiation but was associated with some alterations in expression of genes underlying the specificity of ZNF787 in the CoDiNA network. These results strongly suggest ZNF787 as one repressor of neuronal features. Importantly, the experiment demonstrated that  predictions found by CoDiNA could be experimentally verified. 
	
	\section*{Example applications for situation of more than two networks}
	
	Here we present  two example applications of the CoDiNA method. In the first example, we use CoDiNA to analyse two tuberculosis (TB) studies with patients with and without human immunodeficiency virus (HIV) infection. The second showcase uses data from a study with patients with three types of glioma.  
	
	\section*{CoDiNA applied to a tuberculosis and HIV study}\label{hiv}
	
	\subsection*{Database}\label{database}
	The dataset contains expression data from periferal blood of children and adults from two TB and HIV studies. 
	In this application, our aim is to identify similarities and differences in  TB and HIV in both age groups.
	Both studies are available at GEO; the first one (GSE39941\cite{anderson2014diagnosis}) contains gene expression data from 192 children with TB from Kenya, South Africa and from Malawi (HIV$^+$ $n = 69$, HIV$^-$ $n = 123$); the second one (GSE37250\cite{kaforou2013detection}) contains expression data from 197 adults with TB from South Africa and Malawi (HIV$^+$ $n = 99$, HIV$^-$ $n = 98$).
	Both studies aimed to define transcriptional signatures for detection of TB in patients with and without HIV.
	We used the raw data provided at GEO\cite{edgar2002gene, barrett2012ncbi}, pre-processed and normalised them and performed quality control using \texttt{R} and the \texttt{R} package \texttt{lumi}\cite{lumi, lumi2, lumi3}.
	
	\subsection*{Building the networks}\label{building-the-networks}
	Networks were generated separately for adults and children that are HIV$^+$ or HIV$^-$ using the weighted Topological Overlap (wTO) method for positive and negative interactions\cite{nowick2009differences, gysi2017wto} for all the $13,817$ genes with $1,000$ bootstraps. wTO values $\omega_{i,j}$ that were not significant ($p$-value $\leqslant 0.001$) or with $|\omega_{i,j}| < 0.33$ were set to zero.
	Finding a large absolute wTO-value for a pair of genes means that the expression patterns of both genes are strongly (positively or negatively) correlated. The \texttt{R} package \texttt{wTO}\cite{wTOPackage, gysi2017wto} was used for this calculation.
	The parameters used to build the networks were Pearson product-moment correlation coefficient and bootstrap re-sampling method. 
	
	\subsection*{Defining the Gene vs Disorder Enrichment} \label{defining-the-gene-vs-disorder-enrichment}
	In order to define the disorders enriched in each $\tilde\Phi$ category, we test if the amount of genes associated to HIV or TB classified in each category is different than random using an exact Fisher's test and a proportion test. We combine both $p$-values using the Fisher's method and use this as a weight to filter the results.
	The list of genes associated with HIV Infections, AIDS (Acquired Immunodeficiency Syndrome) or sAIDS (Simian Acquired Immunodeficiency Syndrome) or Tuberculosis was retrieved using the tool Gene 2 Disease tool (GS2D) \cite{fontaine2016gene}. For our disease enrichment analyses we only considered those genes that were measured in the final CoDiNA networks.
	
	\subsection*{Comparing the Tuberculosis networks using CoDiNA} \label{comparing-the-tuberculosis-networks-using-CoDiNA}
	
	This dataset gave rise to three main comparisons of interest: (i) Adults with TB: HIV$^-$ vs HIV$^+$; (ii) Children with TB: HIV$^-$ vs HIV$^+$; (iii) Adults and Children with TB: HIV$^-$ vs HIV$^+$.
	
	In order to associate the $\tilde\Phi$ links to each one of the genes, we used the CoDiNA networks filtered for values where the ratio of the $\Delta_{\widetilde\Phi}$ and $\Delta_{\widetilde\rho}$  is greater than unity. This means that we only present links that are the most distant to the centre:  links with highest scores of being \textit{highly specific}, \textit{highly different} or \textit{highly common}, and are  most well clustered. After assigning the gene $\tilde\Phi$ category, we performed an enrichment test for disorders among high scoring nodes.
	
	\subsection*{Comparing the HIV$^-$ vs HIV$^+$ networks} \label{hiv--vs-hiv-in-adults-no-stretching}
\begin{figure}
		\centering
		\includegraphics[width=1\textwidth]{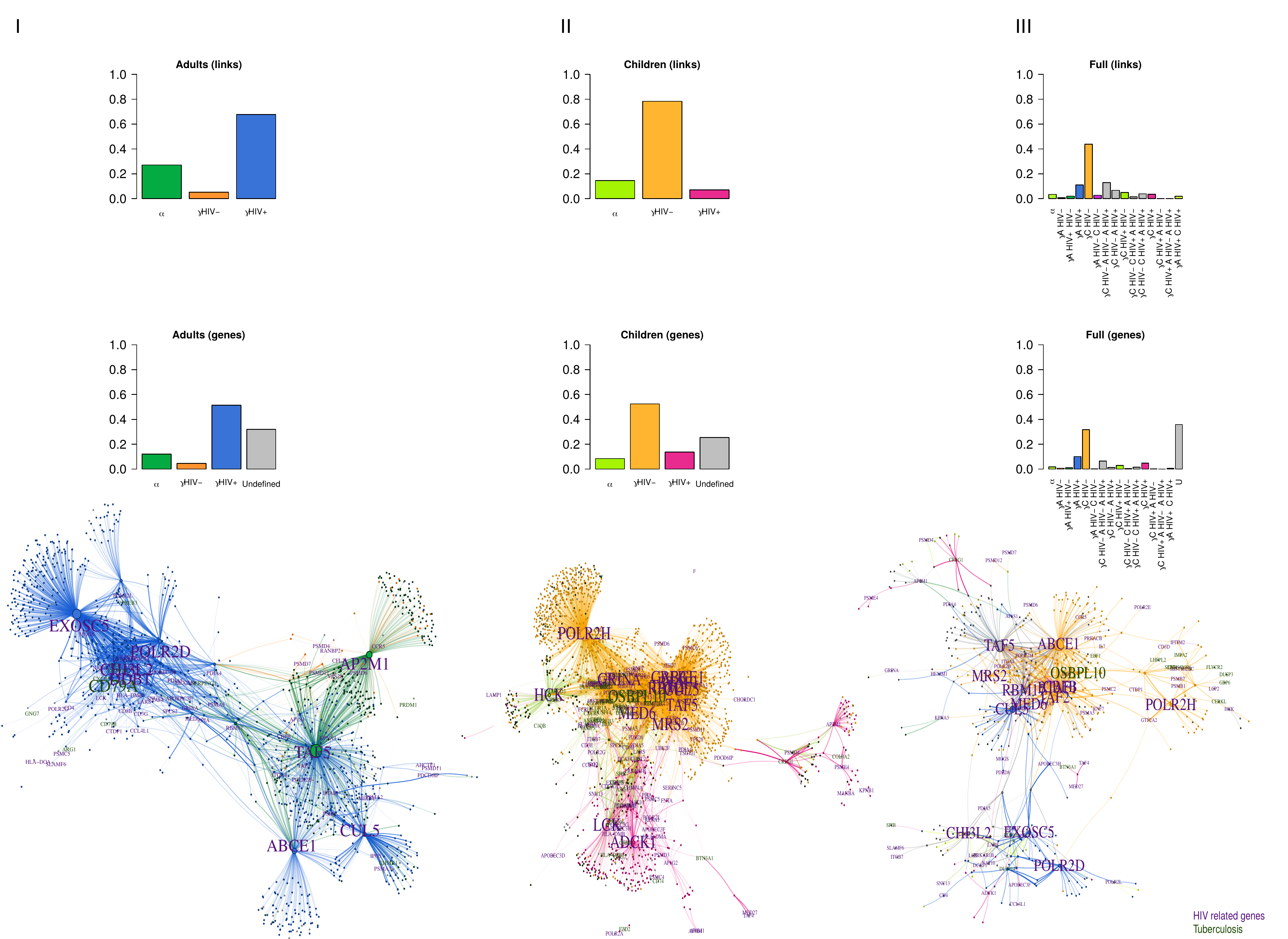}
		\caption{\textbf{Comparing children and adults with tuberculosis and tested for HIV:} The three panels show the categories of links and nodes for each CoDiNA networks as well as the final CoDiNA network for HIV related genes for \textbf{Panel I} adults, \textbf{Panel II} children, and \textbf{Panel III} adults and children. Note that, in the adults network (\textbf{Panel I}) we mainly see genes that are related to HIV$^+$ state, while for children (\textbf{Panel II}), there is a big loss of gene functions as judged from the specificity of genes and links to HIV$^-$ state. This is combined in the full CoDiNA network of adults and children (\textbf{Panel III}).}
		\label{FigHIV}
	\end{figure}

	We compared first the full gene co-expression network of HIV$^-$ and HIV$^+$ adults. 
	In this comparison, CoDiNA was able to identify $80,509$ links and  $3,786$ nodes. 
	From those nodes,  $455$ are $\alpha$ type,  $172$ $\gamma_{_{\text{HIV}^{-}}}$ and $1,948$ $\gamma_{_{\text{HIV}^{+}}}$ (Figure \ref{FigHIV}, panel I), while the remaining nodes were not classified into any of these categories. 
	There were no $\beta$ nodes, although $\beta$ links existed. 
	Importantly, among the nodes not classified as $\alpha$, $\beta$ or $\gamma$, our enrichment analysis showed an over-representation of TB. 
	This is to be expected since all individuals were infected with TB. 
	
	When comparing the networks from the data regarding children, CoDiNA identified $24,3645$ links and  $6,763$ nodes. 
	From those nodes, $573$ are of $\alpha$,  $3,546$ of $\gamma_{_{\text{HIV}^{-}}}$ and $926$ of $\gamma_{_{\text{HIV}^{+}}}$ category, while $1718$ were unclassified (Figure \ref{FigHIV}, panel II). 
	Our enrichment analysis found over-representation of genes related to AIDS and sAIDS for the HIV positive children. The HIV negative group is enriched for genes related to TB. 
	
	Our last comparison included  data from both  children and adults. This final network identified $35,683$ links connecting $4,254$ nodes. 
	The nodes were classified as $77$ of $\alpha$ type, $44$ specific to adults, $33$ to HIV$^-$ adults and $430$ to HIV$^+$ adults. We found $123$ genes associated to children, $208$ to HIV$^+$ children and $1,351$ to HIV$^-$ children. Only $28$ genes are common between HIV$^+$ in adults and children, and only $11$ to HIV$^-$ of both age groups (Figure \ref{FigHIV}, panel III). 
	Among the unclassified genes, we again find enrichment for TB, similar to the network for adults. We further find an association to AIDS in HIV$^+$ positive children and sAIDS in adults and children infected by HIV (Table \ref{CoDiNA:enrich}).
	
	Thus, CoDiNA was able to successfully identify an enrichment of known genes associated with HIV infections among the specific nodes, providing support for the ability of CoDiNA to retrieve biological meaningful results. Importantly, we were also able to pinpoint modules of genes related to each one of the co-infections. 
	
	\begin{table}
		\centering
		\begingroup\tiny
		\begin{tabular}{lllrrr}
			\hline
			Comparison & Disease & Phi & Expected & Observed & weight \\ 
			\hline
			Adults & Tuberculosis & Undefined &  18 &  12 & 0.00 \\ 
			Adults & HIV Infections & $\alpha$ &  92 &   6 & 0.09 \\ 
			Children & Acquired Immunodeficiency Syndrome & $\gamma_{_{\text{Children}_{\text{HIV}^+}}}$ &  22 &   8 & 0.01 \\ 
			Children & Simian Acquired Immunodeficiency Syndrome & $\gamma_{_{\text{Children}_{\text{HIV}^+}}}$ &  18 &   7 & 0.01 \\ 
			Children & Tuberculosis & $\gamma_{_{\text{Children}_{\text{HIV}^-}}}$ &  76 &  45 & 0.02 \\ 
			Children & HIV Infections & $\alpha$ & 162 &  15 & 0.04 \\ 
			Children & HIV Seropositivity & $\gamma_{_{\text{Children}_{\text{HIV}^+}}}$ &   6 &   3 & 0.05 \\ 
			Complete & Tuberculosis & Undifined &  34 &  17 & 0.02 \\ 
			Complete & Acquired Immunodeficiency Syndrome & $\gamma_{_{\text{Children}_{\text{HIV}^+}}}$ &  10 &   2 & 0.05 \\ 
			Complete & Simian Acquired Immunodeficiency Syndrome & $\gamma_{_{\text{Children}_{\text{HIV}^+}\text{Adult}_{\text{HIV}^+}}}$ &   9 &   1 & 0.08 \\ 
			\hline
		\end{tabular}
		\endgroup
		\caption{Disease Enrichment Analysis for each $\Phi$ group in each CoDiNA.} 
		\label{CoDiNA:enrich}
	\end{table}
	
	\section*{CoDiNA applied to a cancer study}\label{CoDiNA-applied-to-a-cancer-study}
	
	\subsection*{Database}
	For the second showcase of our method, we used the gene-expression data from GSE4290\cite{sun2006neuronal}, a study of  patients with glioma. The dataset contains $157$ brain tumour samples of three types ($26$ astrocytomas, $50$ oligodendrogliomas, and $81$ glioblastomas).
	
	The data was downloaded from the GEO website\cite{edgar2002gene}, pre-processed and normalised by ourselves. The cancer expression profiles were normalised with the controls. We used micro-array data, which was analysed using the \texttt{R} environment \cite{R} and the \texttt{affy} \cite{affy} package from the \texttt{Bioconductor}.
	
	The probe expression levels (RMA expression values) and MAS5 detection $p$-values were computed, and only probesets significantly detected in at least one sample ($p$-value $\leqslant 0.05$) were considered. After quality control and probe normalisation, the probes that were not specific to only one gene were removed. If one gene was bound by more than one probeset, the average expression was computed.
	
	\subsection*{Building the networks}
	Because TF deregulation is central to disease progression\cite{richard13, bhagwat2015targeting} in many disease states, and particularly in cancers, we focused on a comparison of the TF co-expression networks between the three different kinds of tumours. To this end, we calculated the wTO network \cite{ravasz2002hierarchical, carlson2006gene, zhang2005general} of the TFs for each tumour. 
	We computed the wTO network for each cancer dataset and the controls separately using only the set of $3,229$ unique TF symbols from the Gene Regulatory Factors (GRF)-Catalogue\cite{berto2016consensus}, filtered by genes with proteins that also are included in the ENSEMBL protein dataset. 
	
	Links with Benjamini-Hochberg adjusted $p$-values\cite{benjamini1995controlling}  smaller than $0.01$ were kept, and links with larger $p$-values  were set to zero.
	Setting the non-significant wTO values to zero ensures that all nodes that were measured can be present in the final CoDiNA networks instead of being removed in the first step of the approach. 
	
	\subsection*{Defining the Gene vs Disorder Enrichment} \label{defining-the-gene-vs-disorder-enrichmentCANCER}
	To verify the enrichment of disorders in each one of the $\widetilde\Phi$ classes, we test if the amount of genes associated to each one of the gliomas under study is different than random using an exact Fisher's test and a proportion test. Both $p$-values are combined into one using the Fisher's method and the resulting $p$-value was used as a weight to filter the results.
	The association of genes to disorders was retrieved using the tool GS2D \cite{fontaine2016gene}. To perform the enrichment test, we used as background only the genes of that list that were expressed in the samples.
	
	\subsection*{Comparing the Networks}
	In total, the CoDiNA network contains  $2,209$ nodes and $206,856$ links above the score ratio threshold. According to the GS2D tool\cite{fontaine2016gene} (weight $< 0.10$) $8$ TFs are described in literature  to be associated to Astrocytoma, $3$ with Oligodendroglioma and $51$ with Glioblastoma (Table \ref{Tab:Cancer}). In our CoDiNA network we identified one of the $8$ known Astrocytoma TFs to be associated with Astrocytoma. Two of the known Oligodendroglioma and $45$ of the known Glioblastoma TFs were associated with the respective glioma types by CoDiNA, providing strong support for the validity of our comparative network approach. 
	
\begin{figure}
		\centering
		\includegraphics[width=1\textwidth]{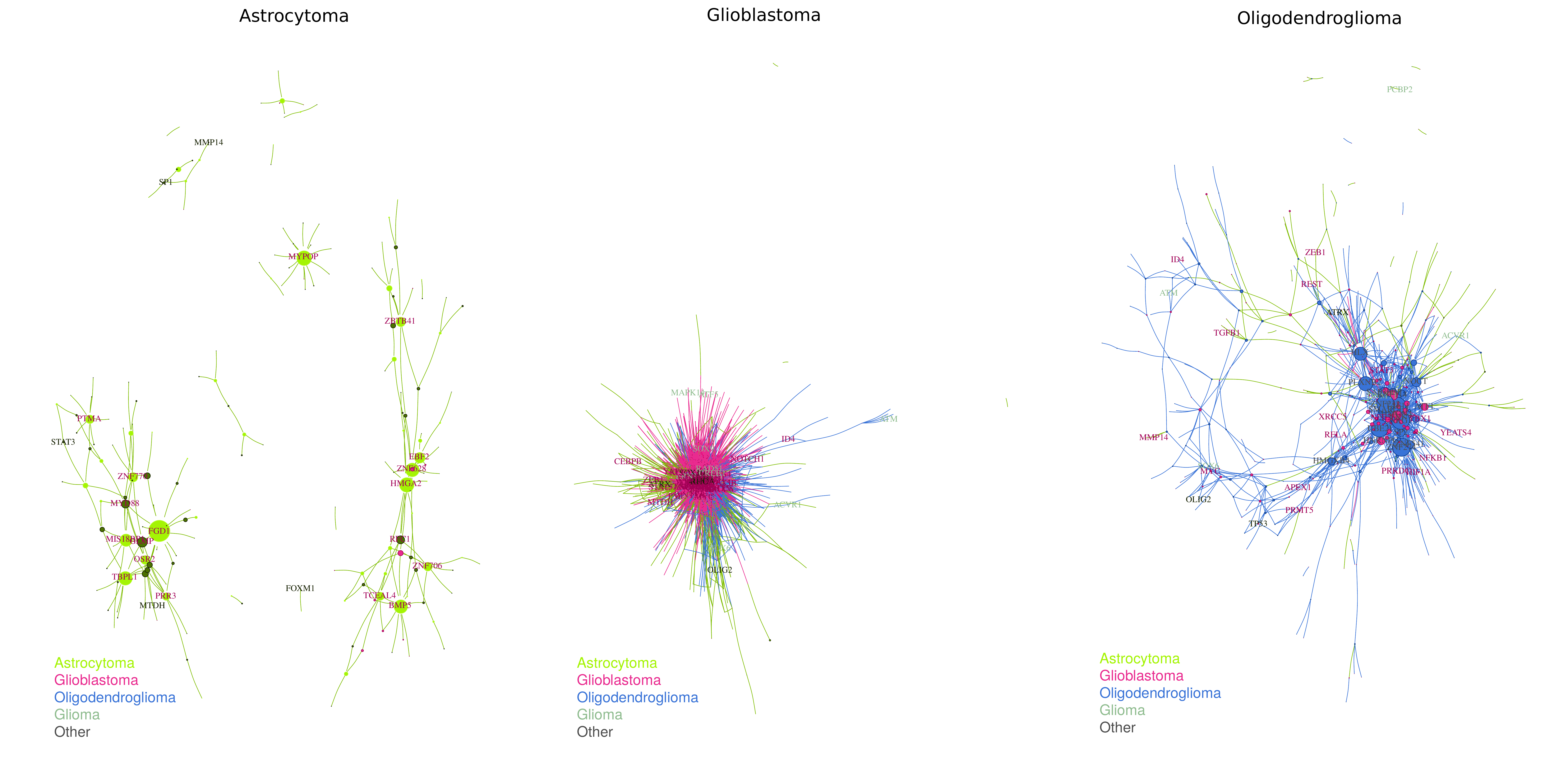}
		\caption{\textbf{TF-TF CoDiNA networks for each of the Glioma types:}  CoDiNA identified TFs with specific co-regulation changes to each cancer, \textbf{Panel I} Astrocytoma, \textbf{Panel II} Glioblastoma, \textbf{Panel III} Oligodendroglioma. \textbf{Panel I}: We can see that mostly glioma and Astrocytoma TFs are present as specific to this network. \textbf{Panel II}:  There is an overlap of TFs involved in other gliomas, but the biggest amount of nodes refers to specific changes in co-expression in Glioblastoma. \textbf{Panel III}: Even though most links and genes are specific to Oligodendroglioma, we can identify overlapping TFs of this cancer and Astrocytoma.}
		\label{Fig:Cancer}
	\end{figure}

	In addition, we identified several TFs specifically associated with Astrocytoma that were not previously linked to this type of cancer (Fig. \ref{Fig:Cancer}, panel I). The TFs with the  $10$ strongest associations are: FGD1, TCEAL4, ZNF628, TBPL1, BMP5, MYPOP, HMGA2, PRR3, MIS18BP1, BMP7. Among these, HMGA2, TBPL1, BMP5 and BMP7 were previously described as associated to neoplasm and neoplasm metastasis\cite{fontaine2016gene}. 
	The most strongly differentiated TFs associated with Oligodendroglioma not previously described (Fig. \ref{Fig:Cancer}, panel II) were: SMARCE1, ZNF274, NRG1, ZNF232, UBE2I, TXK, TAF11, PLXNB2, HLX and SAP30BP. Of these, SMARCE1, NRG1, PLXNB2 and UBE2I were previously described as associated to Neoplasm Invasiveness and neoplasic cellular transformation\cite{fontaine2016gene}.
	The TFs most specific for Glioblastoma (not previously described) (Fig. \ref{Fig:Cancer}, panel III) were: ZNF558, PTBP1, XRN2, RNF114, ZNF45, ZNRD1, KHDRBS2, RFXANK, NIFK and ZNF540. Here, the genes PTBP1, RNF114, XRN2 and  ZNRD1 are described as associated with other neoplams\cite{fontaine2016gene}. This suggests, that CoDiNA can be applied to detect novel candidate genes for specific cancer types. We were able to identify TFs previously associated with other neoplams, but not the types of glioma under study, in important roles in the differential glioma network, indicating that those TFs are also deregulated in those disorders.
	
	\begin{table}
		\centering
		\begingroup\tiny
		\begin{tabular}{llrrr}
			\hline
			Disease & Phi & Expected & Observed & weight \\ 
			\hline
			Astrocytoma & $\gamma_{\text{Astrocytoma}}$ &   8 & 1 & 0.03 \\ 
			Oligodendroglioma & $\gamma_{\text{Oligodendroglioma}}$ &   3 & 2 & 0.06 \\ 
			Glioblastoma & $\gamma_{\text{Glioma}}$ &  51 & 45 & 0.10 \\ 
			\hline
		\end{tabular}
		\endgroup
		\caption{Disease Enrichment Analysis for each $\widetilde\Phi$ group in a Cancer-Glioma Study.} 
		\label{Tab:Cancer}
	\end{table}
	
	\section*{Conclusion}
	We presented a novel method that allows for the systematic comparison of multi-dimensional data in different conditions and the representation of the analysis as a single network. In particular, our method identities links and nodes that are common to all networks under consideration, specific to at least one network, or have different signs among the compared networks. 
	To evalute CoDiNA, we applied it to a neurogenesis study where it identified genes for neuron differentiation, of which which one was experimentally modified, confirming our assessment of it's importance.
	Applying our method to an HIV and TB study, CoDiNA retrieved networks that were enriched for genes involved in HIV. With multiple glioma-type cancer datasets, we identified network signatures that are specific to each type of glioma. More importantly, we were able to identify  genes previously associated with disorders and identify new genes interacting with those. This suggests that our method produces biologically meaningful results. We expect that our method will be helpful for many diverse studies comparing network data generated from multiple conditions, such as different diseases, tissues, species or experimental treatments.
	
	\section*{Availability}
	\texttt{CoDiNA} is open source and freely available from \texttt{CRAN} \url{https://cran.r-project.org/web/packages/CoDiNA/} under the GPL-2 Open Source License. It is platform independent.
	
\section*{Methods}
\subsection*{How the CoDiNA method works}

Let $\mathbb{W}$ be a set of networks constructed using a similarity score, $\rho_{i,j}$ with scalar range $\left[-1,1\right]$, e.g. correlation methods, and each network contains only links with weights found to be statistically significant given a predefined threshold value. For each link, we classify its presence in $\mathbb{W}$ into one of the three following $\Phi$ categories based on its link weights (one link-weight value per network): a link is said to be an $\alpha$ link, if it is present in all networks with the same sign on its weight, i.e., this link is \textit{common} to all networks.  A link is considered as $\beta$ type if it is present in all networks but with \textit{different} signs of the link weights, i.e., the links represents a qualitatively different kind of interaction in at least one network. This interaction might indicate, e.g. that a particular gene changed its function. Finally, a link is considered to be of $\gamma$ type if it is present (i.e.  statistically significant non-zero weight) in some networks but not all, regardless of the sign of the link weight. The $\gamma$ link-type is \textit{specific} to at least one network. 

We may further sub-divide the $\Phi$ categorisation of each link into $\widetilde{\Phi}$ according to the networks containing said link. This sub-classification is mainly important for understanding the behaviour of the $\beta$ and $\gamma$ links. 
Since we work with networks typically constructed using correlations (and thus setting the weight of a link as a correlation measure), the link weight value  $\rho_{ij}~\in~\left[-1,1\right]$. We partition, by default, the interval in three equal parts ($\tau = 1/3$), which will be denoted  as corresponding to a positive link, negative link or neutral link. If the compared networks have different link-weight ranges, these may be normalised by using a multiplicative ({\it stretch}) parameter.

In order to avoid false associations, an important step to be aware of is that a {\it node} should be present in all networks; if a node is absent in at least one network, we remove all of its links in the networks where this node is present.
This step is implemented to prevent the erroneous inference that a particular node is associated with a specific condition, when in fact that specific node possibly was not measured in the other conditions. If a link weight is found not to be significant, we assign it a (weight) value of zero, thus allowing all measured nodes to be included in the analysis even when only some of its links are significant.

For a system consisting of a set $\mathbb{N}$ of nodes, where $N$ is the cardinality of $\mathbb{N}$ in the $\mathbb{W}$ networks, let $\rho_{ij,k}$ be a specific link weight between nodes $i$ and $j$ in network $k$.  Each link is categorised as follow
\begin{align} \label{eq:mean}
\widetilde{\rho}_{ij,k}= &
\begin{cases}
-1 & \textrm{if }  \rho_{ij,k} < -\tau\\ 
1 & \textrm{if }  \rho_{ij,k} > \tau\\ 
0 & \textrm{otherwise},
\end{cases}
\end{align}
where  \CatWeightij~ is an integer transformation of the link weight based on the threshold, and $\tau$, in the standard implementation of CoDiNA, is $1/3$, which allows the space to be divided equally in $3$ parts. If a particular link categorical weight ~\CatWeightij~ is zero in all the $\mathbb{W}$ networks, this link is removed from posterior analyses.

After the correlation values are coded into the categorical variables $\widetilde{\rho}$, each link is assigned to an additional group category that shows in which condition the link is present and what its sign is, if present.
This classification step is particularly important for links that are classified as $\beta$ or $\gamma$ type, because it is straightforward to clearly identify in which network(s) the link is specific or different. 
The maximum number of groups is $G_{\max} = (3^{W} -1)$, where $W$ denotes the cardinality or number of networks in the set. Note that, the group for which all category values are equal to zero is removed from  analyses.
Our approach  is to assign an  $\alpha$, $\beta$ or $\gamma$  classification to each  of the links, by defining $\Phi$ as follow

\begin{align*}
\Phi_{ij} =& \begin{cases}
\alpha_{ij} & \textrm{if } \sum_{k = 1}^{W}|\widetilde{\rho}_{ij, k}| = W \wedge   |\sum_{k = 1}^{W}\widetilde{\rho}_{ij, k}| = {\#(W)} 	\\
\beta_{ij} & \textrm{if } \sum_{k = 1}^{W}|\widetilde{\rho}_{ij,k}| = W \wedge  |\sum_{k = 1}^{W}\widetilde{\rho}_{ij, k}|  < {\#(W)} \\
\gamma_{ij} &  \textrm{otherwise}.
\end{cases}
\end{align*}

Each link receives a sub-category, $\widetilde{\Phi}_{ij}$, based on the pattern of networks in which that link exists. 
This makes it more straightforward to interpret the links in each of the categories, and as a result this improves our ability to identify links that are specific to a subset of networks and the subset of networks it has a different behaviour.

To illustrate the concept of sub-category, assume the following $\widetilde\rho$ of a particular link in $3$ networks: Network A value is 1, Network B is 1 and Network C is also 1. Because the value 1 is common in the three networks, this $\Phi$ category is clearly $\alpha$, and no further explanation is needed. Now, take as a second example, Network A has the value 1, Network B, -1, and Network C 1. Its $\Phi$ class is $\beta$, but this class cannot help us understand where the change occurs, therefore, the $\widetilde{\Phi}$ is needed. Its $\widetilde\Phi$ class is $\beta_{\text{Network2}}$. Important to note is that CoDiNA assumes the first network to be the reference network. As a final example, assume that the $\widetilde{\rho}$ weight of the three networks are 0, 1 and 1 for Networks A, B and C, respectively. This link does not occur in network A, so it is a $\gamma$ links, that is specific to networks B and C. But this is not possible to understand only by reading that its category is $\gamma$, therefore, its $\widetilde{\Phi}$ category is $\gamma_{\text{NetworkB.NetworkC}}$.

Let, $\mathbb{E}$, be the set of links and E the cardinality of $\mathbb{E}$. When all $\mathbb{E}$ links are assigned a $\Phi_{ij}$ category and further sub-categorised as $\widetilde{\Phi}_{ij}$, we score the links to identify those that are stronger. For every node $i = 1,2,\ldots,N$, we interpret the array of link weights $\left(\rho_{1i},\ldots,\rho_{Wi}\right)$ as a point in a $W$-dimensional Euclidean space. In particular, as each link weight is bounded, all points are contained in the cube determined by the Cartesian product $\left[-1,1\right]^{W}$.

As such, a link that is closer to the centre of the $W$-dimensional cube is weaker than a link closer to the links. Based on that, the Euclidean distance, $\Delta$, to the origin of the space is calculated for all links $E_{ij}$. 

However,  since links closer to  corners will trivially have a larger $\Delta$ compared to the others, all distances are penalised by the maximum theoretical distance a link can assume in its category.
Consequently, we define a normalised distance, $\Delta_{ij}$ in Equation (\ref{eq.distnorm}) so that it is in the unit interval
\begin{align}
\Delta_{ij} = \sqrt{ \frac{\sum_{k = 1}^{W} \rho_{ij,k}^2}{\sum_{k = 1}^{W} |\widetilde{\rho}_{ij,k}|}}.
\label{eq.distnorm}
\end{align}
We test if some link-clusters are closer to the border of the cube than the others by a regression model, where the distance to the centre is the dependent variable and the categories are the covariates of the model. 
Indeed, if statistical differences between clusters and distance is detected,
we only select those nodes that belong to one particular cluster of links: the cluster that is the furthest away from the centre. Normalising the distance can be a way to overcome this challenge, by use of Equation.~(\ref{normalization})
\begin{align}
\Delta^*_{ij} = \frac{ \Delta_{ij} - \min(\Delta_{ij})}{ \max(\Delta_{ij}) -  \min(\Delta_{ij})}.
\label{normalization}
\end{align}
Three different approaches may be applied  to the normalisation: 
\begin{itemize}
	\item Normalise all the links together: Here, we do not consider if a complete cluster is situated near the surface or closer to the centre of the cube;
	\item Normalise  links according to their $\Phi_{ij}$ and $\widetilde{\Phi}_{i,j}$ class: In this alternative, all the categories are a part of the final output. This means that if one of the $\Phi$ groups lies inside the cube closer to its centre compared to the other $\Phi_{ij}$ categories, it will be possible to see links that belong to this category in the final network.
\end{itemize}

Another important Score calculated by CoDiNA, called internal Score, denoted by $\Delta_{\widetilde{\rho}}$, measures the distance from the link ${ij}$ to the theoretical best well clustered link in that particular $\widetilde{\Phi}$ category. In other words, if a link is considered an $\alpha$ with all positive links, we calculate its distance to the point $\left(1,1,1\right)$. This score allows us to identify links that are most well defined for each $\widetilde{\Phi}$ category.

Because the two scores $\Delta_{\widetilde{\Phi}}$ and $\Delta_{\widetilde{\rho}}$ are highly negative correlated, the ratio between them also gives us a measure of the very best well defined links. For a well defined CoDiNA network, this ratio should be greater or equal than 1. 

Knowing only the links classification is not sufficient to describe a network; we are also interested in the nodes' classification. To define the $\Phi$ category of a particular node, we make a frequency table of how many times each node had a link in each $\Phi$ category and sub-category. We test, using a $\chi^2$ goodness-of-fit test, if the links of a node are distributed equally in all categories. If the null hypothesis is rejected, the $\Phi$-category with the maximum number of  links is assigned to that particular node. Similarly, the same is done for the $\widetilde{\Phi}$.

\begin{algorithm}[h]
	\caption{Description of the RemoveNodes procedure.}
	\begin{algorithmic}[1]
		\Require Set of $\mathbb{N}$ nodes that belongs to the set of $\mathbb{W}$ networks;
		\Ensure Set of common nodes to all networks\;
		\Procedure{RemoveNodes}{$N_1, \cdots, N_{W}$}
		\State Set\_Nodes = $\bigcap_{j = 1}^{W} Nodes_{j}$ 
		\State \textbf{return} Set\_Nodes.
		\EndProcedure
		
	\end{algorithmic}
\end{algorithm} 

\begin{algorithm}[h]
	\caption{Description of the links categorisation algorithm.}
	\begin{algorithmic}[1]
		\Require Set of $\mathbb{W}$ networks with $\mathbb{N}$ nodes ($W \geqslant$ 2; N $ \geqslant$ 2);
		\Ensure Network with links weight categorised into -1, 0 or 1;
		\State Set $\tau>0$;
		\State By default $\tau \leftarrow 1/3$;
		
		\Procedure{AssignClasses}{}
		\For{$\rho~j \leftarrow 1$ \textbf{to} $W$}
		\For{$\rho_{j}$~$i \leftarrow 1$ \textbf{to} $E$}
		\If{$\rho_{ij}k<\tau$}
		\State {\CatWeightij $\leftarrow -1$;}
		\ElsIf{$j>\tau$}
		\State{\CatWeightij $\leftarrow 1$;}
		\Else
		\State{\CatWeightij $\leftarrow 0$;}
		\EndIf
		\EndFor
		\EndFor
		\EndProcedure
		
	\end{algorithmic}
\end{algorithm} 

\begin{algorithm}[h]
	\caption{Description of the $\Phi$ algorithm.}
	\begin{algorithmic}[1]
		\Require Set of $\mathbb{W}$ networks with $\mathbb{N}$ nodes ($W \geqslant$ 2; $ N \geqslant$ 2);
		\Ensure Network with links categorised into $\alpha$, $\beta$ or $\gamma$\;
		\State Set $\tau>0$;
		
		\Procedure{PhiLinks}{}
		\For{$\widetilde{\rho}$~$i \leftarrow 1$ \textbf{to} $E$}
		\For{$\widetilde{\rho_{ij}}~k \leftarrow 1$ \textbf{to} $W$}

		\If{$\sum_j^W{|\text{\CatWeightij}}| = 0$}
		\State{remove link;}
		\ElsIf{$\sum_j^W{|\text{\CatWeightij}}|~ \& \sum_j^W{\text{\CatWeightij}} = W$}
		\State{$\Phi_{ij}$ $\leftarrow$ $\alpha$;
		}
		\ElsIf{$(\sum_j^W{|\text{\CatWeightij}}|~ \& \sum_j^W{\text{\CatWeightj}}) \neq W$}
		\State{$\Phi_{ij}$  $\leftarrow$ $\beta$;}
		\Else
		\State{$\Phi_{ij}$ $\leftarrow$ $\gamma$;}
		\EndIf
		\If {\CatWeightij $= 1$ }
		\State{ Positive$_{ij} \leftarrow$ 1 }
		\ElsIf {\CatWeightij $= -1$ }
		\State{Negative$_{ij} \leftarrow$ 1}
		\EndIf
		\State $\Delta_{ij}^* \leftarrow$ Equation (\ref{eq.distnorm}).
		\EndFor
		\EndFor
		\EndProcedure
	\end{algorithmic}
\end{algorithm}

\begin{algorithm}[h]
	\caption{Description of the node-categorisation algorithm.}
	\begin{algorithmic}[1]
		\Require Set of $\mathbb{N}$ nodes with $\mathbb{E}$ links ($E \geqslant$ 2; N $ \geqslant$ 2);
		\Ensure Node classified as $\alpha$, $\beta$ or $\gamma$ type\;
		
		\Procedure{PhiNodes}{}
		\For{$i \leftarrow 1 \textbf{ to } N$}
		\State {$\Phi_{\alpha}$ = Count $\alpha$;}
		\State {$\Phi_{\beta}$ = Count $\beta$;}
		\State {$\Phi_{\gamma}$ = Count $\gamma$;}
		\State {Test if  $\Phi_{\alpha}$ = $\Phi_{\beta}$ = $\Phi_{\gamma}$}
		\EndFor
		\EndProcedure
	\end{algorithmic}
\end{algorithm}

\renewcommand{\Call}{\textbf{\newline Call: }{\textsc{}}}
\newcommand{\EndCall}{\textbf{\newline End Call}}

\begin{algorithm}[h]
	\caption{Description of the \texttt{CoDiNA} algorithm.}
	\begin{algorithmic}[1]
		\State Call: \textsc{RemoveNodes}
		\State End Call
		\State Call: \textsc{AssignClasses}
		\State End Call
		\State Call: \textsc{Phi Links}
		\State End Call
		\State Call: \textsc{Phi Nodes}
		\State End Call
	\end{algorithmic}
\end{algorithm} 

\newpage
\section*{Implementation: The \texttt{CoDiNA} \texttt{R} package}
To make the proposed methodology publicly available, an \texttt{R} package, called \texttt{CoDiNA}, was developed, where all the presented steps are implemented. The \texttt{R} package also includes an interactive visualisation tool. The functions included in the package are:

\begin{itemize}
	\item  \texttt{normalize}: Normalise a variable according to Equation~(\ref{normalization});
	\item  \texttt{OrderNames}: Reorder the names of the nodes for each link in alphabetical order;
	\item \texttt{MakeDiffNet}: Categorise all the links into \texttt{$\Phi$}, \texttt{$\widetilde\Phi$} and \texttt{Group} the categories. Calculate the  normalised Scores;
	\item  \texttt{plot}: Classifies the nodes into $\Phi$ and $\widetilde{\Phi}$ following a user-defined cutoff for the chosen distance and plots the network in an interactive graph, where nodes and links can be dragged, clicked and chosen according to its group or classification.
	The size of a node is relative to its degree. Nodes and links that belong to the $\alpha$ (\textit{common}) group are coloured in shades of green and have a triangle shape. Nodes belonging to the $\beta$ (\textit{different}) group are coloured in shades of red and have a square shape.  Nodes of the $\gamma$ (\textit{specific}) group are coloured in blue and their shape is a star. Nodes have a category for group and $\Phi$ or $\widetilde{\Phi}$,  according to a $\chi^2$-goodness of fit test as defined above. If a node is group-undetermined it is diamond-shaped, and if a particular node is cluster-undetermined it is grey.
	The user can also choose a layout for the network visualisation from those available in the \texttt{igraph} package\cite{csardi2006igraph}. 
	It is further possible to cluster  nodes, using the parameter  \texttt{MakeGroups}, and the user may select among the following clustering algorithms:  ``walktrap''\cite{pons2006computing},   ``optimal''\cite{brandes2008modularity},   ``spinglass''\cite{reichardt2006statistical, newman2004finding, traag2009community},   ``edge.betweenness''\cite{freeman1978centrality,brandes2001faster},   `fast\_greedy''\cite{clauset2004finding}, ``infomap''\cite{rosvall2007maps, rosvall2009map}, ``louvain''\cite{blondel2008fast}, ``label\_prop''\cite{raghavan2007near}   and   ``leading\_eigen''\cite{newman2006finding}. These algorithms are implemented in the \texttt{igraph} package\cite{csardi2006igraph}; 
	\item The AST data.table contains the nodes and the weighted topological overlap (wTO) of Transcription Factors (TFs), from GSE4290~\cite{sun2006neuronal} for astrocytomas;
	\item The GLI data.table contains the nodes and the wTO of TFs, from GSE4290~\cite{sun2006neuronal} for glioblastomas;
	\item The OLI data.table contains the nodes and the wTO of TFs, from GSE4290~\cite{sun2006neuronal} for oligodendrogliomas;
	\item And the CTR data.table contains the nodes and the wTO of TFs, from GSE4290~\cite{sun2006neuronal} for controls.
\end{itemize}

	\section*{Acknowledgements}
	
	This work was supported partially by a doctoral grant from the Brazilian government's Science without Borders program (GDE $204111/2014-5$). This work was partly developed during a D.M.G. internship at the Almaas Lab and she expresses her gratitude to members of this group for inputs and great discussions.
	The authors would also like to thank Alvaro Perdomo-Sabogal for sharing  the GRF manually curated database. The authors would further like to thank Wesley Bertoli for proofreading the manuscript.
	
	\section*{Author contributions statement}
	
	Conceptualisation, E.A. and D.M.G; 
	Methodology, E.A., D.M.G, K.N. T.M.F.; 
	Software, D.M.G.;
	Validation, D.M.G., K.N. and V.B.; 
	Formal Analysis, D.M.G., K.N. and V.B.; 
	Resources, D.M.G., E.A, K.N., and V.B.;
	Data Curation, D.M.G.; 
	Writing – Original Draft, D.M.G and K.N.; 
	Writing – Review and Editing, all authors; 
	Visualisation, D.M.G.; 
	Supervision, E.A. and K.N.; 
	Funding Acquisition, E.A., K.N., and V.B.
	
	\section*{Additional information}
	
	Correspondence and requests for materials
	should be addressed to D.M.G.~(email: deisy@bioinf.uni-leipzig.de) or K.N~(email: katja.nowick@fu-berlin.de)

\end{document}